\documentclass{article}
\usepackage{arxiv}
\usepackage{algorithm}
\usepackage{algorithmic}
\usepackage[utf8]{inputenc} 
\usepackage[T1]{fontenc}    
\usepackage{hyperref}       
\usepackage{url}            
\usepackage{booktabs}       
\usepackage{amsfonts}       
\usepackage{nicefrac}       
\usepackage{microtype}      
\usepackage{lipsum}		
\usepackage{graphicx}
\usepackage{doi}
\usepackage{pdfpages}
\usepackage{caption}
\usepackage[thinlines]{easytable}
\usepackage[algo2e]{algorithm2e}
\usepackage{dblfloatfix}

\title{hChain: Blockchain Based Large Scale EHR Data Sharing with Enhanced Security and Privacy}

\author{ \href{https://orcid.org/0000-0002-3047-486X}{\includegraphics[scale=0.06]{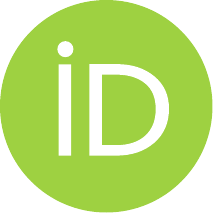}\hspace{1mm}Musharraf N.~Alruwaill} \\
	Department of Computer Science and Engineering \\
	University of North Texas\\
	Denton, TX 76203 \\
	\texttt{MusharrafAlruwaill@my.unt.edu} \\
	\And
	\href{https://orcid.org/0000-0003-2959-6541}{\includegraphics[scale=0.06]{orcid.pdf}\hspace{1mm}Saraju P.~Mohanty} \\
	Department of Computer Science and Engineering \\
University of North Texas\\
Denton, TX 76203 \\
\texttt{saraju.mohanty@unt.edu} \\
		\And
	\href{https://orcid.org/0000-0002-1616-7628}{\includegraphics[scale=0.06]{orcid.pdf}\hspace{1mm}Elias Kougianos} \\
	Department of Electrical Engineering \\
University of North Texas\\
Denton, TX 76203 \\
\texttt{elias.kougianos@unt.edu} \\
}

\renewcommand{\shorttitle}{\textit hChain: Blockchain Based Large Scale EHR Data Sharing with Enhanced Security and Privacy}

\hypersetup{
pdftitle={A template for the arxiv style},
pdfsubject={q-bio.NC, q-bio.QM},
pdfauthor={David S.~Hippocampus, Elias D.~Striatum},
pdfkeywords={First keyword, Second keyword, More},
}

\begin{document}
\maketitle

\begin{abstract}
Concerns regarding privacy and data security in conventional healthcare prompted academics to suggest alternative technologies. In smart healthcare, blockchain technology addresses existing concerns with security, privacy, and electronic healthcare transmission. In addition,  medical data is collected, managed, processed, and stored using  the Internet of Medical Things (IoMT). Integration of Blockchain Technology with the IoMT allows real-time monitoring of protected healthcare data. Utilizing edge devices with IoMT devices is very advantageous for addressing security, computing, and storage challenges. Encryption using symmetric and asymmetric keys is used to conceal sensitive information from unauthorized parties. SHA256 is an algorithm for one-way hashing. It is used to verify that the data has not been altered, since if it had, the hash value would have changed. This article offers a blockchain-based smart healthcare system using IoMT devices for continuous patient monitoring. In addition, it employs edge resources in addition to IoMT devices to have extra computing power and storage to hash and encrypt incoming data before sending it to the blockchain. Symmetric key is utilized to keep the data private even in the blockchain, allowing the patient to safely communicate the data through smart contracts while preventing unauthorized physicians from seeing the data. Through the use of a verification node and blockchain, an asymmetric key is used for the signing and validation of patient data in the healthcare provider system. In addition to other security measures, location-based authentication is recommended to guarantee that data originates from the patient area. Through the edge device, SHA256 is utilized to secure the data's integrity and a secret key is used to maintain its secrecy. The hChain architecture improves the computing power of IoMT environments, the security of EHR sharing through smart contracts, and the privacy and authentication procedures.
\end{abstract}


\section{Introduction}

The rapid growth in population has significantly increased the pressure on healthcare systems, creating a mismatch between the demand for services and the availability of healthcare providers \cite{Everything_smartCity}. Meanwhile, traditional healthcare systems rely on a centralized client-server architecture to store and manage EHRs, resulting in data fragmentation that remains inaccessible to other facilities. This fragmentation places an added burden on patients who receive care from multiple providers, as they must share their updated records from various facilities at each visit. Consequently, a secure solution is required to facilitate seamless EHR sharing \cite{Zhu_Centric_key_management}.

\begin{figure}[htbp]
	\centering
	\includegraphics[width=0.95\textwidth]{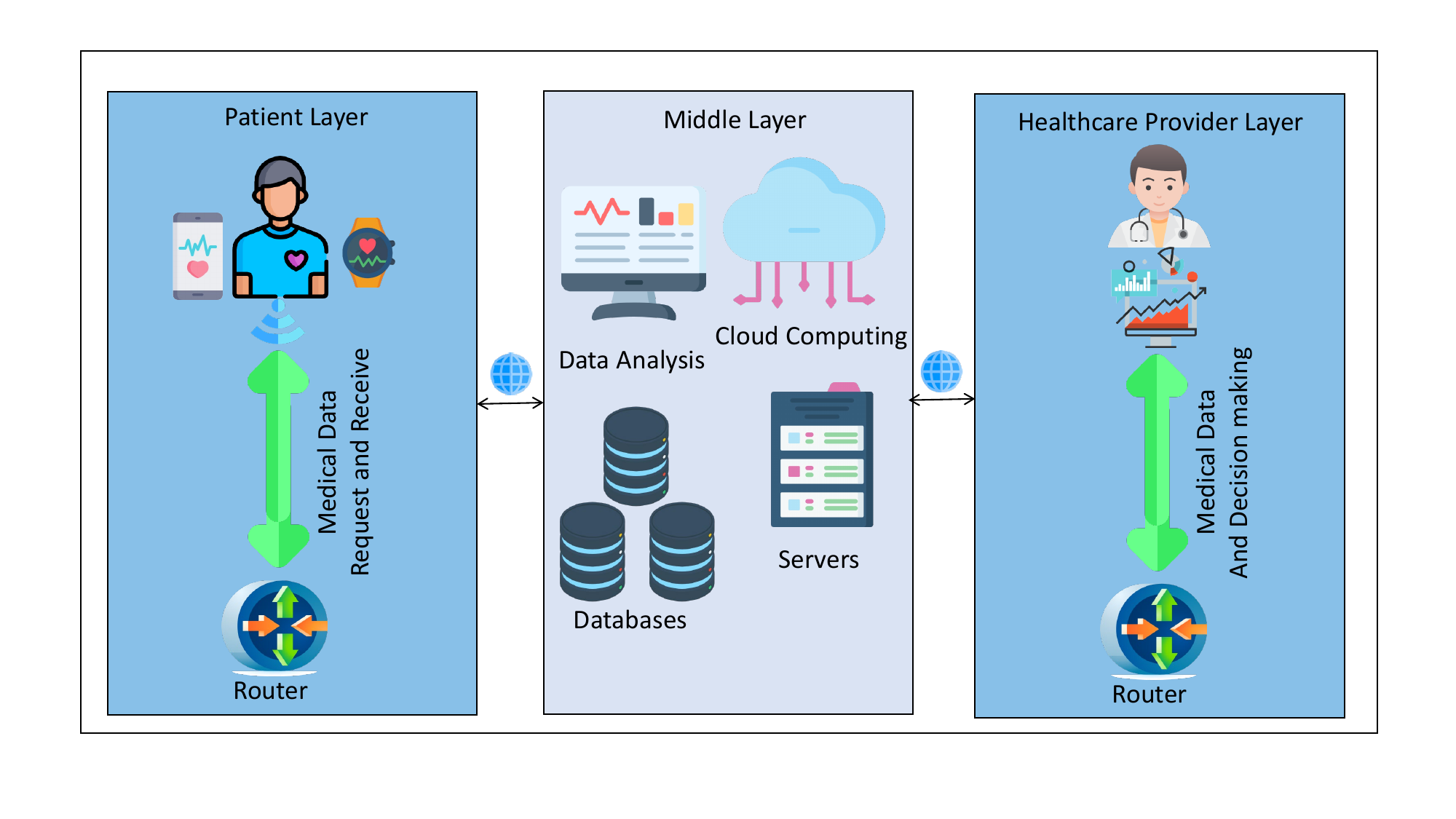}
	\caption{Healthcare Cyber-Physical System Overview.} 
	\label{FIG:System Overviews} 
\end{figure}

Although emerging technologies such as IoMTs and wearable devices have been introduced to enhance patient health through real-time monitoring \cite{RealTime_IoMT}, they often face limitations related to processing power, memory capacity, and security vulnerabilities \cite{survey_Blockchain,IoT_Security_ch}. Moreover, the sensitive nature of medical data heightens these concerns, since any unauthorized access or misuse may seriously compromise patient well-being \cite{iGlu}. As a result, there is a clear need to strengthen privacy measures, enable secure EHR sharing, and maintain patient-centric approaches within a robust security framework.
An overview of a typical healthcare cyber-physical system is shown in Figure \ref{FIG:System Overviews}.

Smart healthcare encompasses an integrated framework of diverse technological solutions and stakeholder collaboration, facilitated by information and communication technologies (ICT) in a smart city context, to ensure the provision of high-quality patient services \cite{Everything_smartCity}. The backbone of smart healthcare is the Internet of Medical Things (IoMT), which enables real-time monitoring of patient health, although it continues to encounter limitations in device performance, data storage capacity, and security measures \cite{Recent_IoMT}. In order to strengthen confidentiality and ensure transparency, blockchain technology offers a decentralized and tamper-resistant approach to securely handling patient records and secure data sharing. In order to strengthen confidentiality and ensure transparency, blockchain technology offers a decentralized and tamper-resistant approach to securely handling patient records and facilitating secure data sharing. These developments underscore the urgent requirement to enhance healthcare systems and integrate advanced technologies that can deliver more efficient, patient-centric care.

The proposed hChain approach utilizes blockchain technology to enhance the healthcare system and smart contracts to support patient-centric and automation processes. It also employs Role-Based Access Control (RBAC) to enable secure data sharing and grant patients greater authority over their medical records. By adopting a decentralized blockchain structure, hChain ensures system availability and reliability. Moreover, location-based authentication strengthens security, and symmetric-key encryption safeguards patient data privacy within the overall framework.

The rest of this paper is organized as follows. 
  Section~\ref{sec:background} provides essential background on smart healthcare. 
  Section~\ref{sec:novel_contributions} outlines the novel contributions of the current work, including the addressed problem and the proposed solution. 
  Section~\ref{sec:related_work} discusses relevant prior research. 
  Section~\ref{sec:hchain_framework} details the hChain proposed solution framework, followed by Section~\ref{sec:hchain_architecture}, which describes the hChain architecture. 
  Section~\ref{sec:hchain_algorithms} then presents the proposed algorithms for hChain. 
  Implementation and experimental results are covered in Section~\ref{sec:implementation_experiments}, 
  and finally, Section~\ref{sec:conclusions_future} concludes the paper and provides suggestions for future research.

\section{Background}
\label{sec:background}

\subsection{Smart Healthcare: Why We Need It}

Healthcare systems are facing mounting pressure due to rapid population growth and the rising prevalence of chronic diseases. These challenges expose the limitations of traditional healthcare models and underscore the need for innovative frameworks to alleviate stress on existing resources. In response, smart healthcare has emerged as a promising paradigm that leverages cutting-edge technologies—such as the Internet of Medical Things (IoMT) devices, wearable sensors, and blockchain—to enable secure, real-time data exchange while maintaining privacy, security, and data integrity. By capturing and analyzing patient health information in real time, healthcare providers can gain timely insights into patients’ conditions, employ artificial intelligence (AI) models to predict and classify potential ailments, and intervene proactively to prevent disease progression rather than respond solely at advanced stages \cite{Intro_Chronic_Prevention}.

\begin{figure}[htbp]
	\centering
	\includegraphics[width=0.99\textwidth]{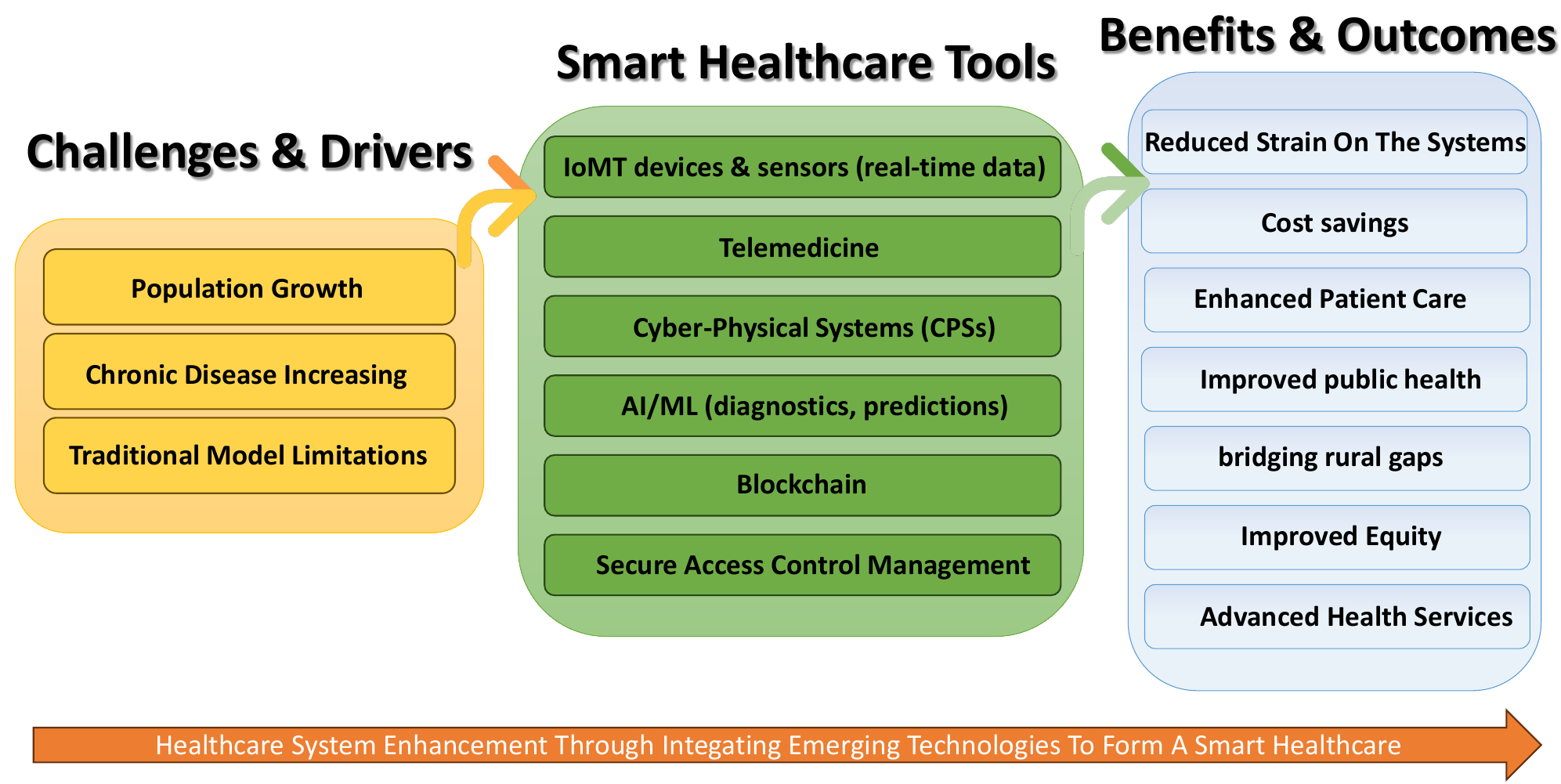}
	\caption{Smart Healthcare Outcomes.} 
	\label{FIG:SH_Importance} 
\end{figure}

Through the integration of IoMT devices and AI, patients benefit from continuous monitoring and personalized treatment plans as depicted at Figure \ref{FIG:SH_Importance}. This proactive approach not only improves patient outcomes but also lowers the incidence of severe chronic diseases \cite{Intro_prevent}, thereby minimizing emergency visits and the overall burden on healthcare providers. Additionally, it can lead to cost savings for patients who are spared from extended hospital stays and expensive treatments. AI-driven models facilitate precise disease detection, classification, and prediction, allowing clinicians to tailor interventions to individual patient profiles, thereby promoting effective, patient-centric care \cite{Intro_Chronic_Prevention}. Thus, smart healthcare benefits significantly from integrating AI, as it enhances health system overall and further strengthens the overall standard of care \cite{SM_AI1,SH_AI2,SH_AI3}.

Beyond clinical advantages, smart healthcare also enhances automation within healthcare systems, diminishing administrative redundancies and operational inefficiencies \cite{Intro_Automation}. By digitizing appointment scheduling and other routine processes, medical staff can devote more time to direct patient care rather than paperwork. This efficient workflow further supports better resource allocation and higher patient satisfaction.

Another critical benefit of smart healthcare lies in telemedicine platforms and remote consultation services, which offer high satisfaction for rural citizens by expanding access to care and reducing barriers \cite{Intro_rural}. By using cloud-based solutions, rural communities gain access to specialized healthcare professionals who may otherwise be geographically inaccessible. Consequently, health disparities between urban and rural populations can be mitigated, improving overall public health outcomes on a broader scale.

Data management is equally vital in this context. The integration of advanced encryption methods, compliance with healthcare regulations, and the utilization of secure blockchain frameworks help safeguard patient information against unauthorized access. These measures encourage greater trust among stakeholders by ensuring that sensitive medical data remains confidential and is accessible only to authorized clients \cite{hchain2}.

In summary, traditional healthcare models struggle to meet the growing demands driven by population increases and the surge of chronic illnesses. Transitioning to smart healthcare can elevate service quality, preserve the sustainability of healthcare systems, and alleviate the pressure on healthcare professionals—particularly when managing urgent and high-risk cases. Moreover, by broadening patient access to specialized care and creating a more equitable healthcare environment, smart healthcare serves as a crucial catalyst for enhancing public health outcomes globally.

\subsection{Smart Healthcare Architecture}

Smart healthcare architecture integrates multiple advanced technologies, including cloud computing, edge computing, and IoMT-based Cyber-Physical Systems (CPS) \cite{Intro_10}. At the foundation is the first layer, composed of IoMT devices and sensors that continuously capture real-time health data. These devices not only monitor patient vitals but can also trigger immediate actions—such as activating robotic aids or other actuators—based on detected abnormalities or threshold breaches. Building on this, the second layer, known as the network and communication layer, ensures secure data transfer across the system. It employs specified protocols and encryption methods designed to maintain confidentiality and integrity as data move between edge devices, local servers, and the cloud.The third layer, referred to as the data management and analytics layer, holds significant importance for various stakeholders. Its primary focus is to store, organize, and process the large volume of data generated in real time, with particular attention to protecting sensitive information. By enabling AI and machine learning frameworks, this layer also facilitates complex data analyses and predictive modeling to inform clinical decisions. The fourth layer, or the application layer, serves as the main interface for clinicians, patients, and administrative staff. Here, stakeholders can visualize health insights, schedule appointments, and provide essential feedback to the system. Finally, an overarching security layer spans all layers of the smart healthcare architecture. It establishes policies, enforces regulations, and implements cybersecurity measures that collectively safeguard the system against potential attacks, ensuring robust data protection throughout every stage of operation. An overview of smart healthcare architecture is shown in Figure \ref{FIG:SH_Arch}. The following sections illustrate each layer of the architecture, including examples of their key functions and components:

\begin{figure}[htbp]
	\centering
	\includegraphics[width=0.95\textwidth]{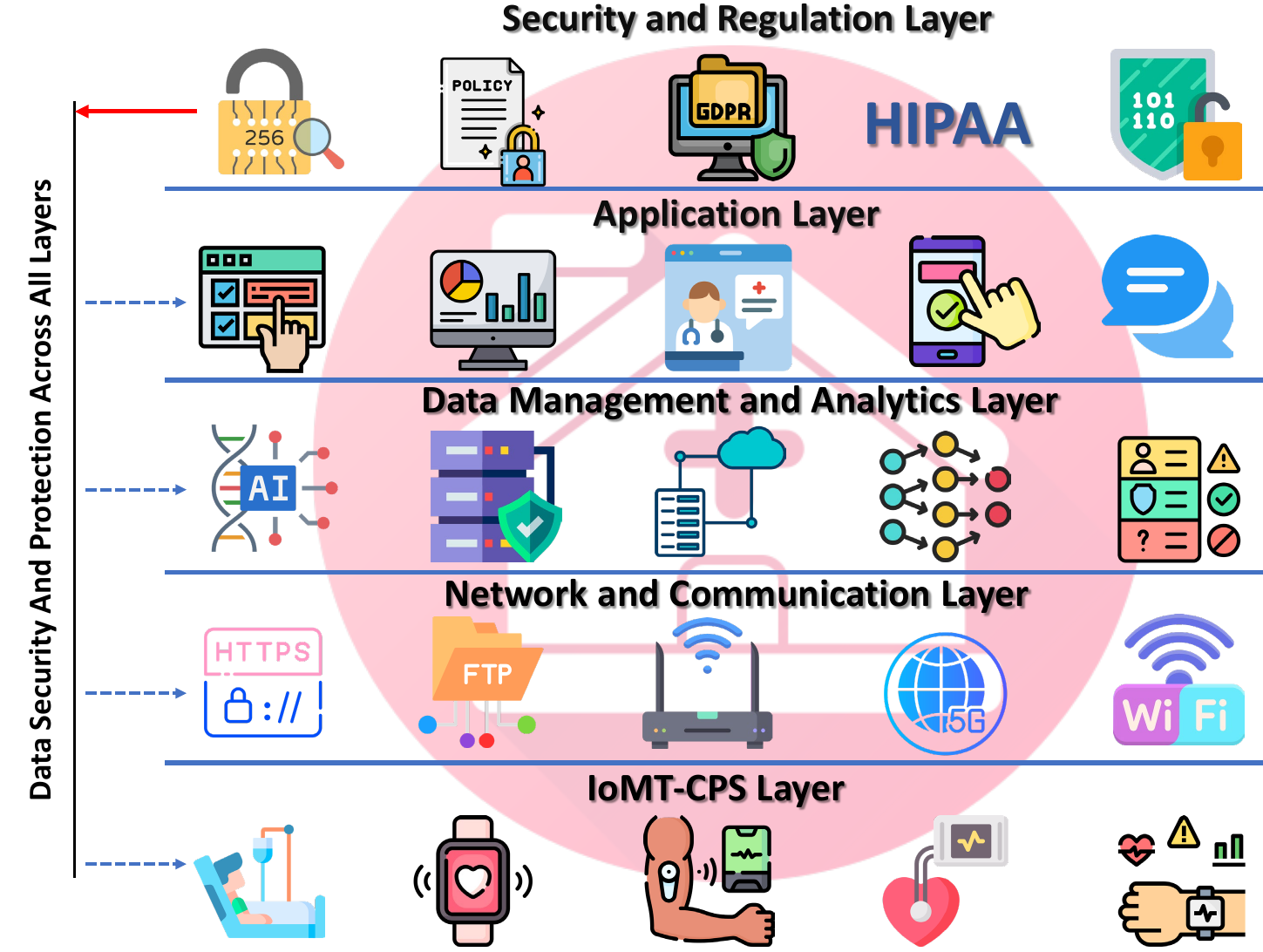}
	\caption{Smart Healthcare Architecture: Layered View.} 
	\label{FIG:SH_Arch} 
\end{figure}

\subsubsection{IoMT–CPS Layer}

This initial layer forms the foundation of smart healthcare architecture, employing IoMT devices, sensors, robotics, and actuators in real time to optimize their combined capabilities—particularly when integrated with AI to enhance patient care. Wearable devices, for example, continuously monitor vital signs and thereby offer deeper clinical insights for more informed decision-making. In addition, these devices can remain on patients throughout their hospital stay, with the CPS responding appropriately to real-time data collected by sensors. For instance, an infusion pump might be linked to ongoing patient monitoring and adjust medication flow rates in response to changes in vital signs.

\subsubsection{Network and Communication Layer}

The networking layer encompasses both communication protocols and the underlying infrastructure to ensure secure data transmission between sensors, edge devices, gateways, and cloud servers. In this context, maintaining confidentiality and data integrity is paramount, allowing information to be stored or retrieved without risk of compromise. Technologies such as Wi-Fi and Bluetooth facilitate device connectivity, while protocols like TCP/IP, HTTPS, and FTP enable robust data exchange. By integrating these protocols, the networking layer upholds the reliability of end-to-end communication throughout the smart healthcare system.

\subsubsection{Data Management and Analytics Layer}

Smart healthcare systems often depend on IoMT-based CPS and produce substantial real-time data volumes. It is therefore crucial to manage and share these large datasets securely, ensuring that only authorized entities can access or store them while maintaining data availability. In practice, healthcare data may be housed on cloud platforms, requiring robust user authentication and authorization mechanisms, along with appropriate key management, to safeguard information integrity. Moreover, the integration of AI within this framework maximizes the value of the collected data by enabling deeper insights into patient health. Through advanced analytics—such as classification models and predictive algorithms—healthcare providers can anticipate chronic disease risks and make evidence-based decisions, ultimately optimizing patient outcomes.

\subsubsection{Application Layer}
This layer employs software tools and web-based applications to facilitate interactions between end-users, such as patients and healthcare providers. Through these interfaces, healthcare services are delivered, patient data can be visualized on dashboards, appointments are scheduled, and secure messaging is enabled. In addition, the system consolidates information from multiple sources into a single, cohesive view, thereby enhancing data accuracy and promoting collaboration across various organizations.

\subsubsection{Security and Regulation Layer}

The security and regulation layers concentrate on enforcing robust data protection measures and ensuring compliance with relevant standards across the entire architecture. These layers define policies governing data ownership and procedural workflows, thereby upholding requirements such as HIPAA or GDPR. They also oversee identity verification, preserve data integrity, and often employ advanced cryptographic methods to counter cybersecurity threats and safeguard the broader system.

\section{Novel Contributions of The Current Paper}
\label{sec:novel_contributions}

At this section, it discuss the problem of the current framework or systems that are currently running around the world and present the disadvantages of healthcare provider systems and security issues as well. It also provides a solution to improve the current system in different aspects and the proposed solution presented as well at the end of this section.

\subsection{Problem Addressed in the Current Paper}
Healthcare providers system currently have many drawbacks that make the system ability to be lose integrity and security and ability to provider more services as well . Centralized system has a security issue and privacy as well \cite{Toward_security}. For example, once the data are stored centric-based, the system may get attack and then lose the data or the data may get changed and ability to provide a services are low comparing to decentralized-based system. IoMT and other emerging technologies are not capable to provide an enough security to the system due to its devices constraints and capabilities. Time, security and services are all has issues in the current system such as time of transferring the patient record from one to another healthcare provider and the patient does not own the record as well so that it takes time to transferring the record due to the verification and validation process. 

\subsection{Proposed Solution}
The centric-based system transferred to decentralized system. It also improve the security of IoMT and wearable devices with the help of advanced security methods. Edge device helps to take over the computation process and solves constraints issues of low capability devices. Blockchain network improve the security of the system and non-repudiation to ensure the confidentiality \cite{Prabha_Telecare}. It also provides access control management through running smart contract and save time and cost to transferring the data from one to another HCP. Authentication improves in multiple factors to ensure the data came from the right source. Validation ensures the data integrity once the data transfer from data source to the destination.

\subsection{The Novelty of the Proposed Solution}
Security and privacy in IoMT and wearable devices are not enough due to their computation power, storage and memory capabilities. IoMT device with the help of near edge device improves the security and time of computation process due to edge has much higher capability than low capability devices.The data may get man in the middle attack during transmission publicly\cite{Authentication_Ehealth}. Therefore, Symmetric key encryption is addressed in this paper to help to hide the data from an authorized person from the data source until stored in blockchain.  Multiple factor of verification and validation improve the whole system security. Blockchain technology stores the data in decentralized manner and process and take an advantages of blockchain  security, integrity and availability to improve the healthcare providers system \cite{Poonguzhali2020}. Authentication based on location proposed in hChain to improve the data integrity and authenticity as well. Smart contract helps to save patient time and improve the security as well. It helps to have ability to the system to provide access control management over the electronic healthcare records. Directory service verifies the user identity internally in healthcare provider before the data being further process to save computation power time and resources and more verification process.

\section{Related Prior Works}
\label{sec:related_work}

Different approaches are proposed by researchers based on various frameworks. For example, they use blockchain based on EHR, Semi-centralized systems to keep store the patient healthcare data and cloud-based data storage. It also differentiates them in which some address using IoMT and some are based on hospital transaction origination instead of home to be the data source. Various approaches have been proposed to secure electronic health records using diverse schemes and specialized mechanisms, each designed to optimize EHR management and overall healthcare system \cite{rw_C1,rw_c2,rw_c3,rw_c4,rw_c5,rw_c6,rw_c7,rw_c8}.

Furthermore, there are existing solutions for tracking and monitoring physiological parameters of the patient or user for various purposes. SaYoPillow continuously tracks the physiological parameters of the person while sleeping in order to capture the correlation between sleep and stress, and then provides the ability to analyze the physiological parameters without the need for an intruder, which is an automated system known as \cite{stress}. The tracking health status system or person behavior tracking is useful not only for health prediction and quality, but it can also be used to prevent adverse events such as \cite{BAC}. It is approached by monitoring the person while driving through physical and psychological behavior to measure the blood alcohol concentration of the driver. However, there are other approaches for smart healthcare, such as \cite{pufchain}, which focuses on hardware security solutions instead of software solutions to enhance system security. Therefore, there are different existing approaches with different purposes.

In \cite{Enhanced_BC_AES}, framework is proposed comprising of a secure database, blockchain for PHR Personal healthcare record and AES for encrypting the data before being stored. However, the system was not addressing more security relating to authentication, IoMT and emerging technology. In addition, verification before transacting is important where our proposed system is verifying the data and identity each time the data being received. The data source is home and using one way hash function to ensure the integrity beside AES to hide the healthcare data. Moreover, hChain proposed smart contract to have the patient ability to share the data with multiple HCPs and uses Authentication based location beside other authentication factors such as private and public key and user unique identity. 

In \cite{biometric}, the cloud is used to store the patient's data and blockchain to store the indexes of EHR that are stored in cloud to solve the privacy and big data issues. In addition, it uses fingerprint hash for patient identification. However, storing data in cloud is subject to have third party to gain access to patients data. Moreover, once the cloud is attacked, then, the data permanently may be removed. hChain uses blockchain technology to store encrypted EHR to ensure the availability and security due to blockchain characteristics as well as healthcare provider framework for data validation before storing the data on blockchain.

In \cite{Overlay_cloud}, a secure system mainly based on overlay network, cloud storage  and smart contracts is proposed. In addition, it supports automated response based on patients wearable device readings and respond based on the agreements on smart contract normal and abnomal readings. Howerver, it keeps the data on  cloud storage and the hash on overlay network which is risky if the cloud get attacked and the cloud storage have all the patients data. Moreover, it uses the public key of the user to encrypt the data which is secure but it consumes time due to computational complex of public key encryption \cite{Symmetric_Asymmetric}. Therefore, hChain uses Symmetric encrpytion to avoid the latency. It also uses blockchain to store the patient data to overcome the shortcomings of cloud storage. In addition, it uses multiple authentication factors to ensure the data is integrity and authenticated before stored in blockchain.

In \cite{Secure_HCP}, the proposed framework consists of two mainly blockchain networks “Personal Healthcare PHC” and “External Record Management ERM”. PHC maintains the patient data that are generated by wearable devices while ERM for maintaining the doctors and other stakeholders data for the patient such as medicine and bills. The system also proposed ML algorithm for abnormality data detection. However, the system didn’t provide advanced and further solutions for data source verification and authentication. In addition, the data need to be hidden and raise the privacy by encrypting the data before storing the data permanently in blockchain. Two blockchain networks make the framework more complex to be integrated. IoT and wearable devices are capability constrains so that it needs to have a solution for security,  privacy  and device to manage and enhance the throughput of devices constraints.

In \cite{BC_EHR_SM}, a framework is proposed using blockchain to access control management of the data and provides data integrity beside hospital local database. The system strengths the data integrity once the hash value of the data in local database, are stored on blockchain. Therefore, the hash value ensures the data are not being changed once the hash value of the data source matches the existing hash value in blockchain. However, the local database is not efficient to be relied on once it is hacked. Availability, confidentiality are on risk when the data only stored on centralized manner. The attacker can destroy the data. The hash value is one way function so that the system cannot reverse it to reveal the data. hChain stores the data permanently on blockchain to overcome availability issue. The data once generated are encrypted so that the data are in form of unreadable format. In addition, the healthcare framework has verification nodes to verify the data integrity as well as the data source using directory service and authentication based on location to have double verification methods. Smart contract provide more flexibility to share the data in secure manner.

In \cite{Centralized_Validation}, the proposed system uses distributed database to maintain the healthcare data instead of using blockchain to have the data being decentralized. The blockchain is addressed at the system only for accessing control management through the smart contract. Healthcare nodes are only can accessing to the data on DDB through smart contract. The accessing control management of the system is robustness via using smart contract. However, distributed database cannot ensure the data integrity and availability. In addition, the system is not proposed to keep tracking patient health data remotely using IoMT while the system are within healthcare provider. hChain uses advanced AES algorithm to encrypt the data and store the data in decentralized storage using blockchain. Moreover, the authentication is based on different factor which are authentication based on location, user identity, healthcare key pairs and verification node to ensure the data is confidential.

In \cite{BC_HCP_EHR_Sharing}, the proposed system provides suitable architecture for ensuring the data integrity to Machine Learning model through blockchain. In addition, the system provides ability to share the patient record using NEM blockchain. It also provides ability for medical researchers to diagnosis the disease while the patient identity is hidden. However, the system has less security concertation to encrypt the data and the system is not proposed any emerging technology to be used which is that great machine learning feeding as it is real time data and huge data will make the machine learning model trained more. Our proposed system ensures the integrity and privacy and authenticity through using symmetric key, user identity validation, one way hash function and authentication based on location. These steps and algorithms are enhancing the whole framework with security, privacy and integrity. hChain proposed a validation node at each healthcare provider to make sure the whole data are came from the right person and the exact data that came from the source. In addition, it uses emerging technologies IoMT to enhance the healthcare services with new technologies.

Table \ref{TAB:Storage} compares this paper to the previous cited works in terms of storage type and format while Table \ref{TAB:Devices} compares the system devices and security scheme.

\begin{table}[htp]
\caption{EHR Storage Type and Format Comparative View}
\centering
\medskip
\begin{TAB}(r,1cm,1cm)[8pt]{|c|c|c|}{|c|c|c|c|c|c|c|c|c|}
	\textbf{The Framework/System}   & \textbf{EHR Data Storage} & \textbf{EHR Storage Format}    \\ 
	Gabriel and Sengottuvelan \cite{Enhanced_BC_AES} &  Blockchain& Encrypted  \\
	Al Baqari and Barka \cite{biometric} & Cloud Storage& Plain-text\\
	Srivastava et al. \cite{Overlay_cloud}&  Cloud Storage& Encrypted \\
	Chakraborty et al. \cite{Secure_HCP}&  Blockchain And Cloud Storage & Plain-Text \\
	V. B. et al. \cite{BC_EHR_SM}& Database & Plain-Text\\
	Simpson et al. \cite{Centralized_Validation}& Database & Plain-Text \\
	Haddad et al. \cite{BC_HCP_EHR_Sharing}& Blockchain And Cloud & Plain-Text\\
	\textbf{Current Work (hChain)} & \textbf{Blockchain} & \textbf{Encrypted}\\
\end{TAB}
\label{TAB:Storage}
\end{table}

\begin{table}[htp]
	\caption{Systems Devices and Security Comparative View}
	\centering
	\medskip
\begin{TAB}(r,1cm,1cm)[6pt]{|c|c|c|c|}{|c|c|c|c|c|c|c|c|c|}
	\textbf{The Framework/System}   &  \textbf{Edge}   & \textbf{Authentication Factor}  & \textbf{Level of Security}   \\ 
	Gabriel and Sengottuvelan \cite{Enhanced_BC_AES} &  No & Single Authentication Factor & One Level of security \\
	Al Baqari and Barka \cite{biometric} &  No& Single Authentication Factor & One Level of security \\
	Srivastava et al. \cite{Overlay_cloud}&   No& Two Authentication Factors & Multiple Layers of Security \\
	Chakraborty et al. \cite{Secure_HCP}&  No & Single Authentication Factor & One Layer of Security \\
	V. B. et al. \cite{BC_EHR_SM}& No & Single Authentication Factor & One Layer of Security\\
	Simpson et al. \cite{Centralized_Validation}&  No & Single Authentication Factor & One Layer of Security\\
	Haddad et al. \cite{BC_HCP_EHR_Sharing}& No & Single Authentication Factor & One Layer of Security\\
	\textbf{Current Work (hChain)} & \textbf{Yes}  & \textbf{Three Authentication Factors}  &\textbf{Multiple Layers of Security}\\
\end{TAB}
\label{TAB:Devices}
\end{table}

\section{A Novel Blockchain and Multi-factor Authentication Based Smart Healthcare Framework for Enhanced Security, Privacy and Authenticity}
\label{sec:hchain_framework}

\begin{figure}[htbp]
	\centering
	\includegraphics[width=0.95\textwidth]{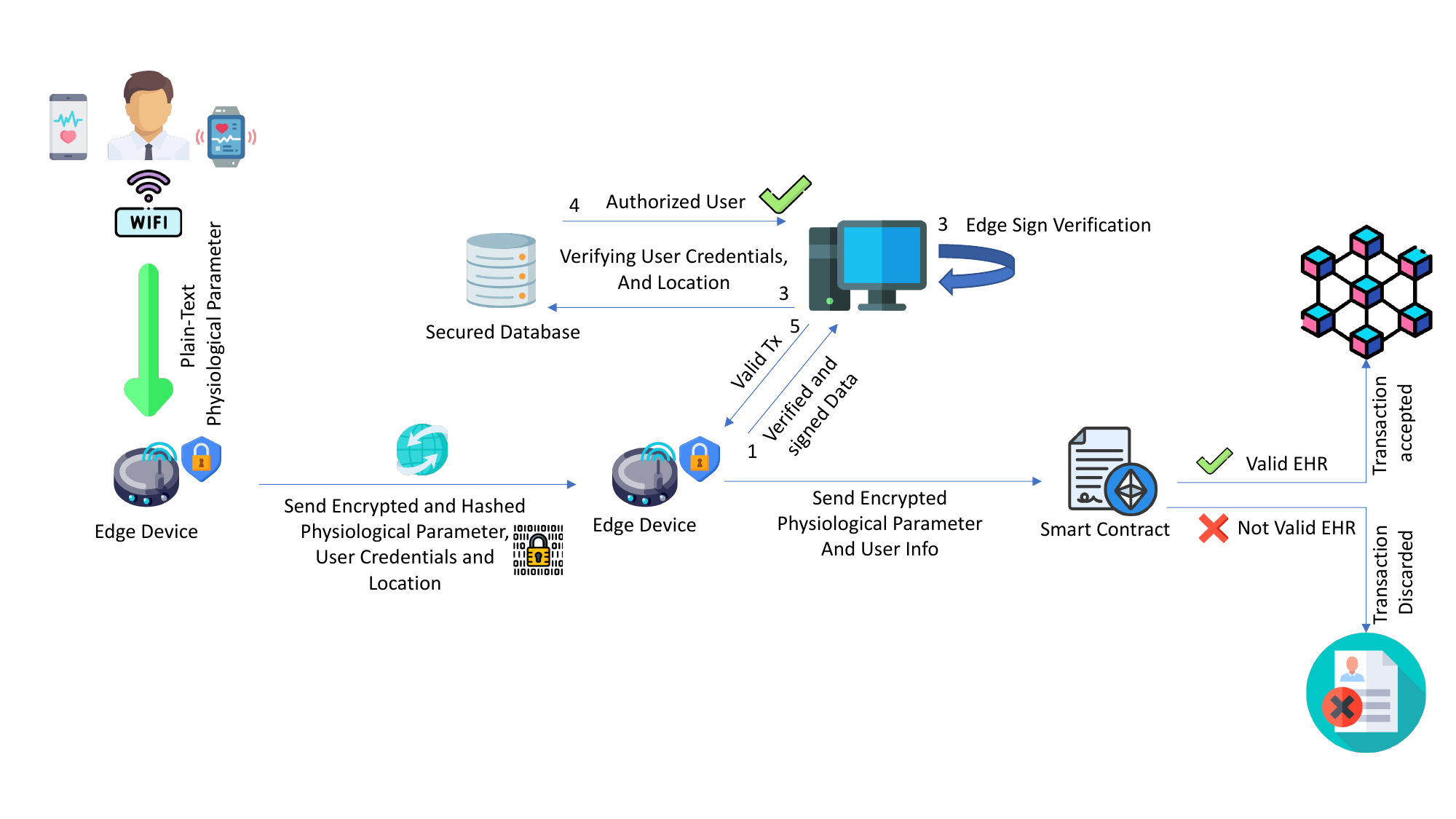}
	\caption{hChain System Overview.} 
	\label{FIG:System Overview} 
\end{figure}

We discuss the proposed healthcare framework components and the novelty of Multi-factor Authentication in this section. The technologies and methods utilized by the framework are also covered. Blockchain-based intelligent healthcare enhances system security and data integrity using blockchain characteristics \cite{Prabha_Telecare}. These characteristics are required for electronic health records EHR to keep data in a secure manner. Multifactor Authentication is proposed for the framework in order to increase the system's authenticity. The framework addresses location-based authentication and user identity in addition to blockchain authenticity via public key infrastructure. Smart healthcare framework components, blockchain technology, location-based authentication, and smart contracts will be discussed in the following subsections. A system overview of hChain is shown in Figure \ref{FIG:System Overview}.

\subsection{Framework Components}

The proposed system is primarily comprised of five components: Blockchain technology, IoMT devices, Verification node, Public and private key encryption, and symmetric key encryption. The components communicate and interact through various mechanism. These are covered in depth below.

\subsubsection{Blockchain Technology}

Blockchain technology can be broadly classified into three categories—private, public, and consortium—depending on the level of openness and access rights \cite{BC_IoT_Challenges}. Structurally, it consists of a sequential chain of blocks, with each block appended in succession \cite{blockchain_Tech_def}. Notably, the inherent properties of blockchain—such as decentralization, encryption-based data protection, transparency, and immutability—further bolster system security and integrity. These features collectively support more robust data management practices and mitigate risks within a wide range of applications.

\subsubsection{Smart Contract}

Smart contracts are self-executing software programs deployed on a blockchain, designed to foster trust among network participants \cite{SmartContract_Security}. By handling incoming transactions from members of the blockchain, a smart contract automates predefined operations and enforces access control policies. Each contract is written and recorded on the blockchain as a transaction, thereby enabling reliable and secure data management. In the hChain framework, the smart contract manages  EHR interactions and ensures that all data exchanges with the blockchain occur in a controlled, authenticated manner.

\subsubsection{IoMT}

IoMT devices collect real-time patient health data through embedded biosensors \cite{IoMT_security}. Although these devices enable remote vital sign monitoring and facilitate immediate clinical analysis, they often operate under resource constraints, offering limited computational power, memory, and storage. Nevertheless, leveraging their continuous data streams can significantly improve both short-term clinical decisions and long-term understanding of patient health. To address the shortcomings inherent in such low-capability devices, hChain integrates a nearby edge device that offloads more demanding tasks, thus ensuring efficient data processing and enhancing overall system performance.

\subsubsection{Edge Device}

Edge devices are positioned in close proximity to IoMT and wearable devices, thereby mitigating the limitations associated with low-capacity hardware. By providing additional storage, memory, and computing power, these edge nodes assume tasks that would otherwise be challenging for resource-constrained devices \cite{edge_ioT_ref,edge_IoT}. In the proposed hChain framework, the edge device first validates incoming data to ensure its integrity and privacy, then forwards it to the verification node for subsequent authentication.

\subsubsection{Verification Node}

The verification node resides within the healthcare provider's infrastructure and is responsible for authenticating user identity, validating data integrity, and confirming key ownership. This process encompasses two factors: first, the verification of user credentials via cryptographic keys, and second, the validation of the user's location through embedded GPS data. Moreover, the verification node verifies each transmission step by hashing and comparing the data to detect any potential modifications. Finally, it communicates with the smart contract to either add or retrieve information on the blockchain.

\subsubsection{Symmetric And Asymmetric  Key Encryption}

Symmetric key encryption, frequently termed a “secret key” method, relies on a single shared key for both encrypting and decrypting data. In contrast, asymmetric encryption employs a pair of keys (public and private) to achieve the same objective \cite{symmetric_asymmetric2}. Given the reduced computational overhead of symmetric encryption compared to asymmetric approaches, it is adopted here to minimize latency \cite{Symmetric_Asymmetric}. Because IoMT devices generate plaintext data—making it vulnerable to unauthorized interception over the internet—symmetric key cryptography is used to enhance both authenticity and confidentiality. These cryptographic processes are executed at the home edge device, thereby offloading the task from the resource-constrained IoMT unit.

\subsection{Location Based Authentication}

The hChain framework employs GPS coordinates as part of its multi-factor authentication strategy. Upon completion of patient registration, the patient's home coordinates are recorded in a secured internal database accessible only to healthcare provider nodes. These coordinates remain in an encrypted format to preserve confidentiality, ensuring that even if an unauthorized party were to obtain the encrypted data, they would be unable to interpret it. Subsequently, when new coordinates are received, the edge device at the healthcare provider layer compares the incoming data to the stored reference. If the calculated distance falls within the authorized home region, the data is considered valid; otherwise, the system rejects the data and terminates the process.

\subsection{Smart Contract For Access Control Management}

Each patient retains an EHR, accessible solely by that individual or an approved entity (e.g., a physician). A smart contract is used to manage access control, enabling users to grant read permissions to specific providers. This mechanism is governed by RBAC, which enforces varied authorization levels according to assigned roles. For instance, each patient maintains a HashMap containing unique provider identities; once permissions are granted, a provider can view the data until the patient modifies or revokes that authorization.

The smart contract organizes users into three membership categories. The first, Administration, has the authority to create or revoke other memberships, thus holding overarching control within the system. The second, HCP Registration, is tasked with adding new HCP memberships, often for broad regional coverage and centralized enrollment. Lastly, HCP itself registers patients on the blockchain and maintains their data once the provider verification process is complete (see Section 4). Together, these membership types enable efficient onboarding of new providers and support flexible access management within the hChain environment.

\begin{figure}[htbp]
	\centering
	\includegraphics[width=0.99\textwidth]{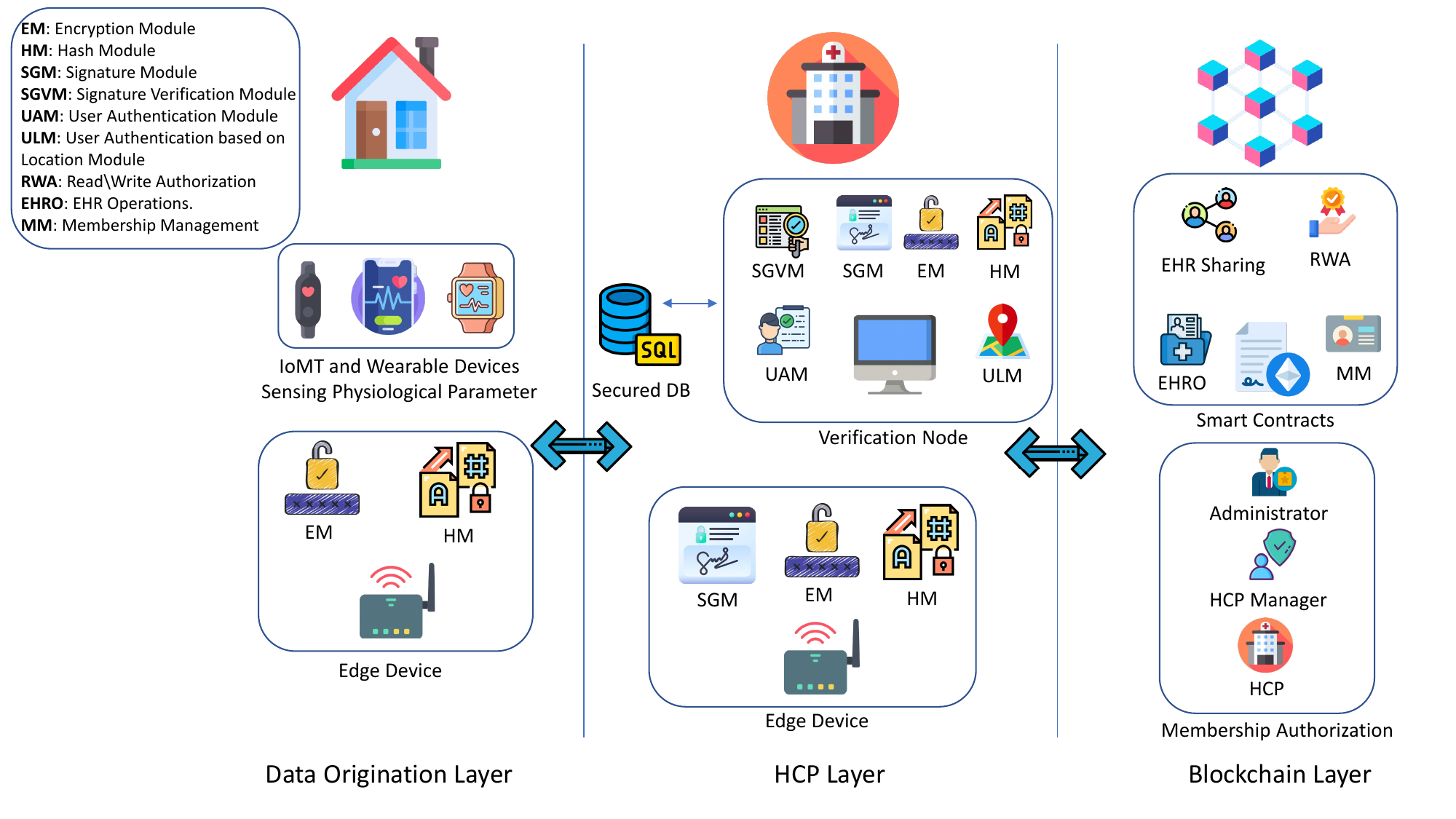}
	\caption{hChain System Architecture.}
	\label{FIG:System Architecture}
\end{figure}

\section{hChain Architecture}
\label{sec:hchain_architecture}

hChain is divided into four primary layers, each responsible for distinct operations and security processes. The first is the IoMT layer, which incorporates IoMT and wearable devices alongside a home edge device to strengthen security and mitigate the limitations of low-capability units. The second layer, representing the healthcare provider’s framework, hosts the HCP edge node and the verification node. The blockchain, situated at the fourth layer, is a private network featuring three key memberships (HCP, HCP Registration, and Administration), as depicted in Figure~\ref{FIG:System Architecture}. Figure~\ref{FIG:Seq} illustrates the interactions among the system’s components in a sequence diagram, showing the data flow step by step. The subsections that follow describe each layer of hChain in further detail.

\subsection{Data Origination Layer}

\begin{figure}[htbp]
	\centering
	\includegraphics[width=0.99\textwidth]{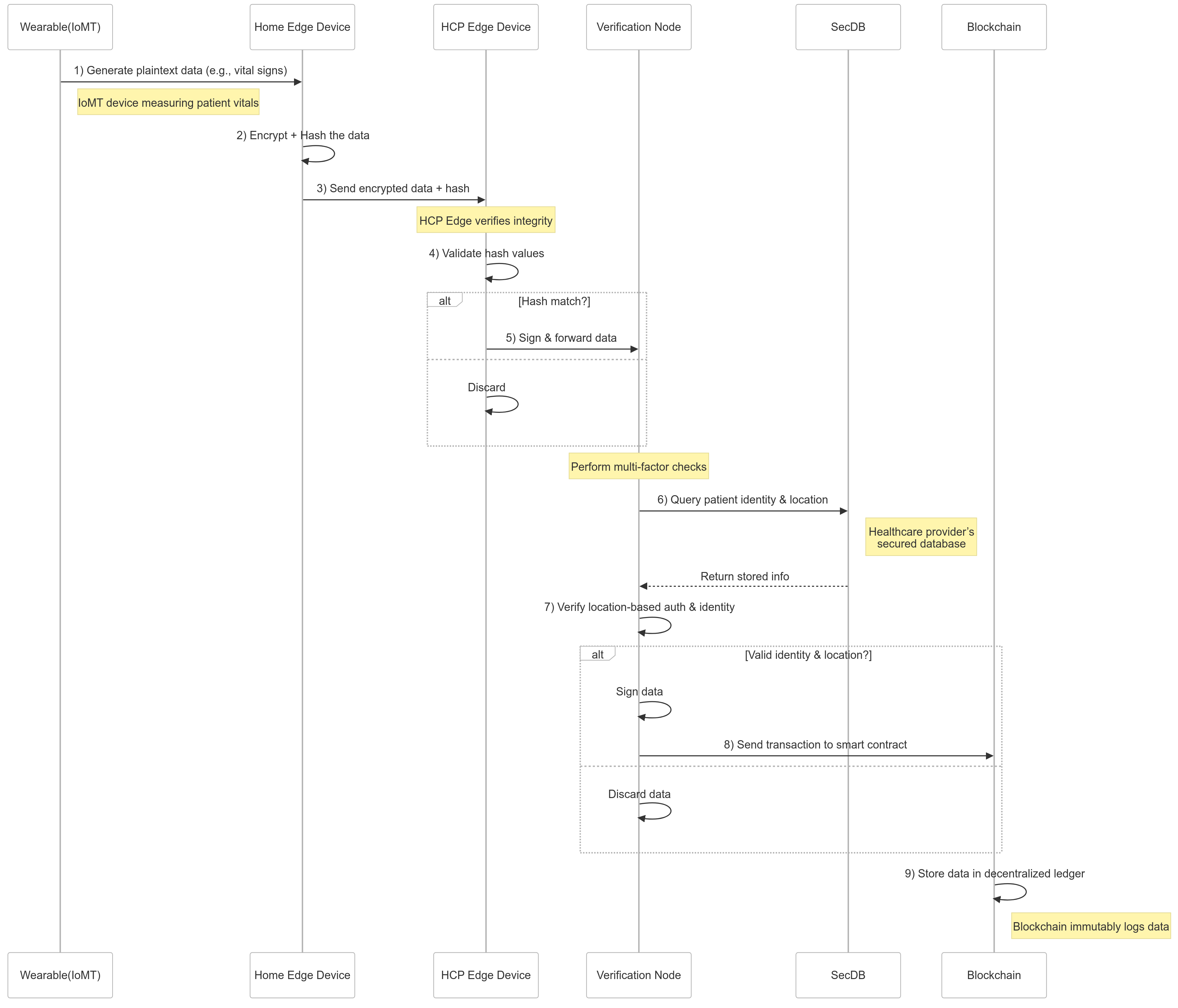}
	\caption{Sequence Diagram of System Interaction.} 
	\label{FIG:Seq} 
\end{figure}

This layer forms the initial stage of the hChain framework, where data originate from patient-worn sensors. These measurements constitute the fundamental input to hChain and typically consist of physiological parameters. Because the edge device provides higher computational capacity, the raw data are initially transmitted in plaintext to the home edge device, making encryption at the IoMT level unnecessary. Data transfer occurs via Zigbee or Wi-Fi (see Figure~\ref{FIG:System Architecture} and Figure~\ref{FIG:Seq}). The edge device stores the secret key and is thus responsible for encryption, hashing, and JSON-encoding the payloads. After encryption using a secret key and hashing with SHA256, the data are encapsulated in JSON for consistent processing at both the HCP layer and the blockchain layer. A secure communication channel then relays these encrypted payloads to the second layer.

\subsection{HCP Layer}

Located in the second tier of the hChain framework, the HCP layer comprises three main entities responsible for verifying data before publication on the blockchain. The first entity is the HCP edge device, which directly receives data from the home edge device. It validates integrity by comparing hash values; if a match is confirmed, the data are forwarded to the verification node. Otherwise, the data are discarded to preserve system resources. Equipped with a key pair, the HCP edge device signs each valid data group, ensuring that the verification node recognizes the source as legitimate.

The second entity is the verification node, which carries its own key pair and handles tasks such as user identity verification, location-based authentication, and signature validation. Upon receiving data from the HCP edge device, the verification node confirms the device’s signature. If the data pass verification, the node extracts the patient’s coordinates from the data group and queries the secured database for reference. This database, forming the third entity within the HCP layer, stores user identities, coordinates, and other essential patient information. A distance calculation then assesses whether the incoming data match the correct patient’s location. If validated, the node proceeds with further authentication and authorization, retrieving the patient identity from the database as well. An unregistered or invalid identity prompts immediate rejection. Once all checks are passed, the verification node signs the data and initiates a transaction on the blockchain via the smart contract, which includes multiple functions to manage data requests. Ultimately, the data are permanently recorded in the blockchain.

\subsection{Blockchain Layer}

The blockchain layer fulfills two primary objectives: storing data in a secure, transparent manner  and delivering robust access control throughout the framework. Access management is governed by a smart contract implementing RBAC, ensuring that only authorized personnel can view or modify specific resources. hChain’s blockchain relies on Ethereum, where data blocks are linked sequentially to form a decentralized ledger \cite{bcdef}. This decentralized approach enhances security, while the use of Proof of Stake (PoS) as a consensus algorithm mitigates the high resource consumption typically associated with Proof of Work (PoW) \cite{PoW_PoS}. 

\section{The Proposed Algorithms For hChain}
\label{sec:hchain_algorithms}

To facilitate direct IoMT or wearable device connections with a healthcare provider, the patient must first be enrolled in the provider's Hashmap. Algorithm \ref{ALG:DataGeneratingAlgorithm} describes every stage of data handling, from initial collection until it arrives at the healthcare provider's edge device. At the beginning, biosensors measure the patient's physiological parameters and transmit them as plaintext to the home edge device. Because hChain does not rely on low-capability devices for computationally intensive operations, a factor that would otherwise introduce latency, data processing—including encryption and structural organization—occurs on the edge device, referred to as PED in Algorithm \ref{ALG:DataGeneratingAlgorithm}. Upon receiving the plaintext, PED encrypts it using a locally stored secret key, thereby mitigating exposure to man-in-the-middle attacks and ensuring that only the holder of the same key can decrypt the data. Then, the encrypted data is grouped and hashed, and this hash value is appended for organizational and integrity verification. The data set also includes GPS coordinates, patient identifiers, and physiological measurements. Once the HCP edge device, known as HCP-E, obtains the grouped data, it checks for integrity by applying the same hashing method. If the hash does not match, the data is discarded; otherwise, HCP-E signs the data using its private key so the verification node can confirm the source and validity. Finally, HCP-E sends the signed data to the verification node, as detailed in Algorithm \ref{ALG:DataGeneratingAlgorithm}.

\begin{algorithm}[htbp]
	\caption{The Steps of Sending Medical data From IoMT to HCP.}
	\label{ALG:DataGeneratingAlgorithm}	
	\small	
	\DontPrintSemicolon
	\SetAlgoLined
	\SetKwInOut{KInput}{Input}
	\SetKwInOut{KOutput}{Output}
	\SetKwInOut{KOterms}{Terms}
	\KInput{\textit{Data $D_i$ collected from IoMT Devices}}
	\KOutput{\textit{Authenticated $GPD$ or Discarded $GPD$}}
	\KOterms{\textit{$UI$ User Unique Identification, $PED$ Patient Edge Device, $HCP-E$ Healthcare Provider Edge Device, $PD$ Patient Medical Data, $GPD$ Group of Patient data, $VN$ Verification Node }}
	\tcc{IoMT Sensing and Generating $PD$}\;
	IoMT Device sends plain-text $PD$ in real-time to $PED$
	
	$PED$ Receive $PD$ and Encrypt $PD$ using Secret Key \;

	$PED$ Hashes Each $PD$ and Append it to $GPD$ \;

	$PED$ Encrypts $UI$ and append it to $GPD$. \;

	$PED$ Groups all $PD$s and $UI$ as $GPD$.
	
	$PED$ sends $GPD$ to $HCP-E$.
	
	$HCP-E$ receives $GPD$s.
	
	$HCP-E$ generates hashes of all encrypted $PD$ in $GPD$.
	
	$HCP-E$ generates hash of $GPD$.
	
	\tcc{$GPD$ consists of multiple of $PD$s}\;
	
	\eIf{ Generated Hashes of $HCP-E$    MATCH    $PD$ hashes in $GPD$ AND Generated Hash of $GPD$ MATCH Appended Hash of $GPD$
	}{ 	
			The generated hash value matched $PD$ hashed value.
			
			$HCP-E$ uses own private key to sign the $GPD$ as signed $GPD$. \;
			
			Append the signature to signed $GPD$. \;
			
			$HCP-E$ sends signed $GPD$ to $VN$.
			
	}{
		DISCARD
	}
	
\end{algorithm}

Signature validation, patient authentication through identity checks and location-based data, and transaction verification are managed by the verification node (VN). Upon receiving data from HCP-E, the VN confirms the HCP-E signature. Once the signature is deemed valid, the VN authenticates the patient's identity by comparing the encrypted patient credentials in a secure database with the encrypted identity transmitted, as outlined in Algorithm \ref{ALG:SendingData}. Subsequently, the VN employs a two-factor approach, with location-based authentication serving as the second method. This involves extracting the patient's GPS coordinates from the transmitted set of data and matching them against corresponding values stored in a secure database. Calculating the distance between the reference coordinates and the patient's location further substantiates authenticity. After all data have been validated and verified, the VN submits the data to the smart contract as a transaction.

\begin{algorithm}[htbp]
	\caption{The Steps of Sending Medical data From IoMT to HCP.}
	\label{ALG:SendingData}	
	\small	
	\DontPrintSemicolon
	\SetAlgoLined
	\SetKwInOut{KInput}{Input}
	\SetKwInOut{KOutput}{Output}
	\SetKwInOut{KOterms}{Terms}
	\KInput{\textit{Verified $GPD$}}
	\KOutput{\textit{Store $GPD$ in Blockchain}}
	\KOterms{\textit{$UI$ User Unique Identification, $HCP-E$ Healthcare Provider Edge Device, $GPD$ Group of Patient data, $VN$ Verification Node. $SM$ Smart Contract, $BC$ Blockchain }}
	
	$VN$ Receive verified and signed $GPD$ from $HCP-E$. \;
	
	$VN$ verifies $HCP-E$ Signature. \;

	\eIf{ Signature is Valid
	}{ 	
		
		$VN$ extracts the $UI$ location from $GPD$. \;
		
		$VN$ sends a $UI$ location query to secured DB. \;
		
		Secured DB responses with stored $UI$ location to $VN$. \;
		
		$VN$ Receives $UI$ Location. ;\
		
		$VN$ Calculates the distance between $UI$ location in $GPD$ and stored $UI$ location. \;
		
		\tcc{Location MUST belong in range of $UI$ Location. }\;
		\eIf{$UI$ Location is Valid} {
			Authentricated \;
			
			$VN$ verifies the patient identity.\;
			
			$VN$ extracts the patient identity from the $GPD$.\;
			
			$VN$ sends a query to Secured DB with patient identity.\;
			
			Secured DB finds the patient identity and return the result to $VN$.\;
			
			\eIf{Patient Identity is Valid} {
			Patient is authenticated and authorized.\;
			
			$VN$ signs $GPD$ with private key. \;
			
			$VN$ Transacts signed $GPD$ to $SM$. \;
			
			$SM$ has Hash Map of Each HCP privileges and Hash Map of each $UI$ belongs to HCP. \;

			$SM$ verifies HCP privileges. \;
			\eIf{HCP writes privilege is Valid} {
				\eIf{$UI$ is Registred} {
					Append $GPD$ to $UI$ Hash Map.
				}{Error message appear to request $UI$ Registeration.\;}

			}{Error Message appear and DISCARD the process.\;}
		}{DISCARD}
		}{DISCARD}
	}{
		DISCARD
	}
	
\end{algorithm}

\section{A Specific Implementation}
\label{sec:implementation_experiments}

Multiple methods and technologies are employed to implement hChain. As illustrated in Figure \ref{FIG:System Overview}, the framework integrates blockchain, smart contracts, encrypted databases, and cryptographic algorithms. Ganache serves as a private blockchain environment, effectively simulating the Ethereum blockchain~\cite{Quick_blockchain_book}, while Brownie is used to compile, deploy, and test smart contracts~\cite{Brownie}. To protect data privacy and authenticity, the system relies primarily on symmetric and asymmetric cryptography. In addition, SQLite stores patient data within the healthcare provider’s infrastructure, thereby managing identity information. Python, a high-level programming language~\cite{Python}, operates alongside Brownie to communicate with the Ganache blockchain and supports command-line functionality for edge devices and the Verification Node.

\begin{figure*}[htbp]
	\centering
	{\includegraphics[width=0.70\textwidth]{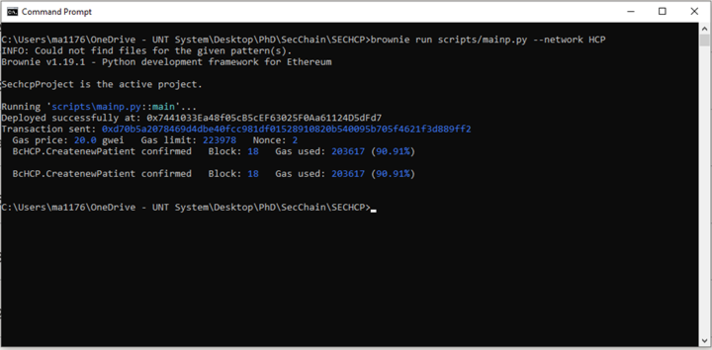}}
	{\includegraphics[width=0.70\textwidth]{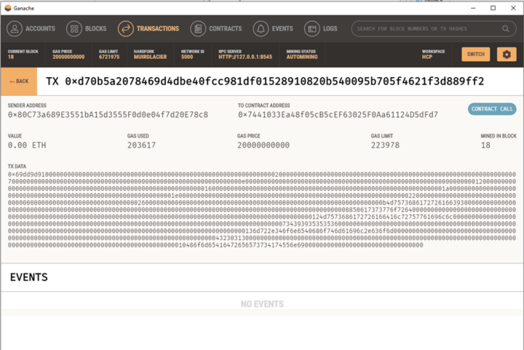}}
	\caption{hChain Smart Contract Implementation }
	\label{FIG:DeplymentAndEnrollment}   
\end{figure*}

Figure \ref{FIG:DeplymentAndEnrollment} illustrates the newly added data on the blockchain as well as the corresponding transaction recorded in Ganache. The data flow starts from the patient side and continues until it is stored on the hChain network. Moreover, the smart contract is deployed on the Sepolia test network. Ethereum is crucial in covering the deployment costs of the smart contract, which is essential for testing and research. When pushing the contract to the testnet, Metamask is used to facilitate payments, this wallet streamlines interactions with the Ethereum blockchain. Accordingly, Ganache is employed as the local blockchain for initial development, while the smart contract is later deployed to Sepolia. Etherscan further offers a web-based interface to track transaction details validated on the Ethereum blockchain.

All implementation steps were performed on a workstation powered by an Intel\textsuperscript{\textregistered} Core\textsuperscript{TM} i7-10700 CPU at 2.90\,GHz and 16\,GB of RAM. This setup provided a stable environment for local testing and subsequent deployment.

\section{Experimental Results}
\label{sec:experimental}

In this section, we examine the security analysis at multiple operational layers. The hChain architecture highlights both the time sensitivity of real-time data and the required level of protection at each stage. In particular, the encryption time analysis compares the use of blockchain-based public–private key cryptography with the proposed symmetric key scheme in hChain, illustrating the trade-offs in latency and security strength.

\subsection{Security Analysis}

The security analysis of hChain consists of three main security principles. The first security aspect involves multiple authentication processes at different layers of the hChain framework. The second principle concerns authorization to access blockchain data through smart contracts and the HCP layer. The third relates to data integrity and data quality—essential for research purposes. Collectively, these principles provide a high level of security for the end user, namely the patient, helping to avoid unauthorized access and ensuring data tampering resistance.

\subsubsection{Unauthorized Access: Transmission and Storage}

There are two primary scenarios in which unauthorized access can occur: during data transmission and once the data is stored. The first type of unauthorized access arises if data is intercepted in transit via a man-in-the-middle attack, allowing an attacker to obtain a copy of the information being sent. Within hChain, symmetric key encryption is employed to encrypt data prior to transmission. This particular method is chosen due to its capability to handle large volumes of text and data, contrasting with the limitations of asymmetric key encryption (detailed in the subsection on encryption time analysis). As a result, the transmitted data remains unreadable and is forwarded to the healthcare provider layer in encrypted form. Even if an attacker manages to capture it, the content is still encrypted and cannot be deciphered without the relevant key. Therefore, hChain addresses man-in-the-middle threats through modern encryption algorithms coupled with secure communication channels.

The second type of unauthorized access occurs when data at rest is targeted. The proposed hChain system implements multiple levels of authentication to protect stored data. Access to the data requires progressing through various authentication steps. The first step involves matching the user’s identity with a secured database maintained within the healthcare provider’s infrastructure. Because the identity is both stored and transmitted in encrypted form, it is difficult for an attacker to interpret the raw data. Further, the data itself comprises multiple segments, obscuring any direct knowledge of which portion contains the user identity, since these segments are sent as part of a collective dataset. The second authentication process employs location-based verification. In this scheme, coordinates indicate where the data is physically associated. After a patient registers with the healthcare provider, the patient’s location is recorded in encrypted format. Subsequently, authorization for accessing data mandates sending the patient’s current GPS coordinates in encrypted form to the healthcare provider system. Once these coordinates are received, the healthcare provider’s edge node calculates the distance and verifies whether the source of the request is situated within a predefined home area. Should the data source lie outside the authorized zone, the user cannot proceed to the next stage of authentication. Consequently, an attacker would need multiple elements to gain full data access: the encryption key employed for data security, valid proof of being at the correct home location, the patient’s unencrypted identity, and the keys for the healthcare provider’s edge node. This means that an attacker would have to compromise not only the patient’s home devices but also the entire provider ecosystem, including edge nodes, the verification node, and the secured database—an endeavor that is considerably complex, as each layer implements distinct security measures.

\subsubsection{Authorization and Brute Force Attacks}

Authorization through brute force attack prevention is accounted for within hChain. A brute force attack typically tries to guess a password or user identity, often by employing wordlists or randomly generated strings. hChain’s multi-layer authentication across different locations and utilizing various keys makes brute force attempts far less feasible. One step involves verifying the patient’s identity. Additional elements include the patient’s location, the patient’s cryptographic key, and keys used by the edge device and verification node. Collectively, these components underlie the authentication process.

When a patient submits their identity, it includes the patient’s current location. These data are encrypted using the patient’s secret key, followed by a hash function to preserve data integrity. The healthcare provider nodes thereby ensure the data remains unchanged en route. The patient’s home system stores only the identity and location, whereas the provider’s infrastructure retains edge device and verification node keys. Thus, each phase of the process unfolds in succession. Even if an attacker successfully brute forces the patient’s identity, they still cannot obtain authorization to access the patient’s records due to the necessity of multiple additional keys, including those for the edge node, verification node, and the patient’s own secret key. Further, any brute force attempt must begin with acquiring the patient’s secret key in the first place; otherwise, it becomes impossible to progress toward retrieving healthcare data. This multi-key, multi-location framework complicates brute force efforts, as no single piece of information suffices to unlock the entire system.

\subsubsection{Data Tampering Prevention}

Data tampering proof is provided in hChain via a one-way hash function. During transmission, such as when sending physiological readings or other healthcare data, the information is first encrypted and then hashed before traversing the internet. Consequently, any alteration—whether during a man-in-the-middle attack or if the edge node itself were compromised—would change the overall hash value \cite{hashfunction}. For example, an attacker might capture the transmitted data, modify the physiological parameters, and relay them to the healthcare provider, but the resulting mismatch in hash values would be detected by the edge node, prompting it to discard the data entirely. This mechanism effectively bolsters data integrity by leveraging the one-way hash function. The same hash function also serves to confirm authenticity: as soon as the data is hashed, all associated documents likewise become hashed, guaranteeing that the receiving side obtains an unaltered file, thereby preserving both its source origin and integrity.

\subsubsection{Collision and Preimage Attack Prevention}

Moreover, hChain addresses collision and preimage attacks. Collision attacks occur when an adversary discovers two distinct inputs that yield an identical hash result, enabling them to create phony data while undermining authenticity \cite{collision_attack}. Preimage attacks focus on a hash output, seeking to identify the specific input that can generate that targeted output \cite{preimage_def}. Within hChain, the deployment of SHA-256 thwarts both approaches, as this hash function is resistant to collision and preimage exploits. By incorporating SHA-256, the system elevates data reliability and integrity, a requirement for medical environments that handle valuable patient data and depend on precise research findings.

\subsection{hChain Input And Output Analysis}
\label{sec:layer}

As presented in Table \ref{TAB:experimental}, there are three level of operations which are patient layer, HCP layer and Blockchain Layer. Patient layer security relies on encrypting the data through the patient edge device using symmetric key encryption. Therefore, the input is plain-text and the output is encrypted data. Moreover, it uses hash function to hash the data to ensure the integrity. Therefore, at the patient layer solve issues related to data integrity and confidentiality.  It also avoid revealing the data from the man in the middle attack. At HCP layer, the data is verified integrity through edge device and sign the data to acknowledge the verification node the data is valid. After that, the verification node verify patient identity through patient username and password. Additionally, verification nodes authenticate the patient based on location. Therefore, at HCP layer verifies patient identity and authenticity for authorization purpose. Moreover, it verifies the signature to ensure it belongs to edge device signature. Hence, it avoids any possibility of modifying the data during transmission and unauthorized access via three authentication factors. At blockchain layer, it verifies the patient if it is matched and registered or not. Once, all previous steps are valid, then, the data is stored in the blockchain.

\begin{table*}[htbp]
	\caption{hChain Experimental Result}
	\centering
	\medskip
	\begin{TAB}(r,1cm,2cm)[2pt]{|c|c|c|c|c|}{|c|c|c|c|c|c|}
		\textbf{Task} &  \textbf{Time (sec)} & \textbf{Input}  & \textbf{Output} & \textbf{Layer of Operation}   \\ 
		\textbf{Data Encryption}& 0.0030083656311035156 & \shortstack{Plain-Text \\Physiological \\ Parameter} &  \shortstack{Encrypted \\Physiological\\Parameter}  & Patient Layer \\
		\textbf{User Identification} & 0.015624046325683594  & Username Identification & \shortstack{Authorized or \\Not Authorized} & HCP Layer \\
		\textbf{Signature Validation}& 0.0009999275207519531 & HCP Edge Signature  & Valid or Not Valid& HCP Layer\\
		 \textbf{\shortstack{Authentication \\ Based Location}} & 0.015622377395629883 & GPS Location &  \shortstack{Authenticated or \\Not Authenticated} & HCP Layer  \\
		\textbf{EHR Tx} &  6.777848958969116 & Encrypted Data & Stored in Blockchain  & Blockchain Layer   \\
	\end{TAB}
	\label{TAB:experimental}
\end{table*}

\subsection{Encryption Time Analysis}
\label{sec:time}

\begin{figure*}[htbp]
	\centering
	{\includegraphics[width=0.80\textwidth]{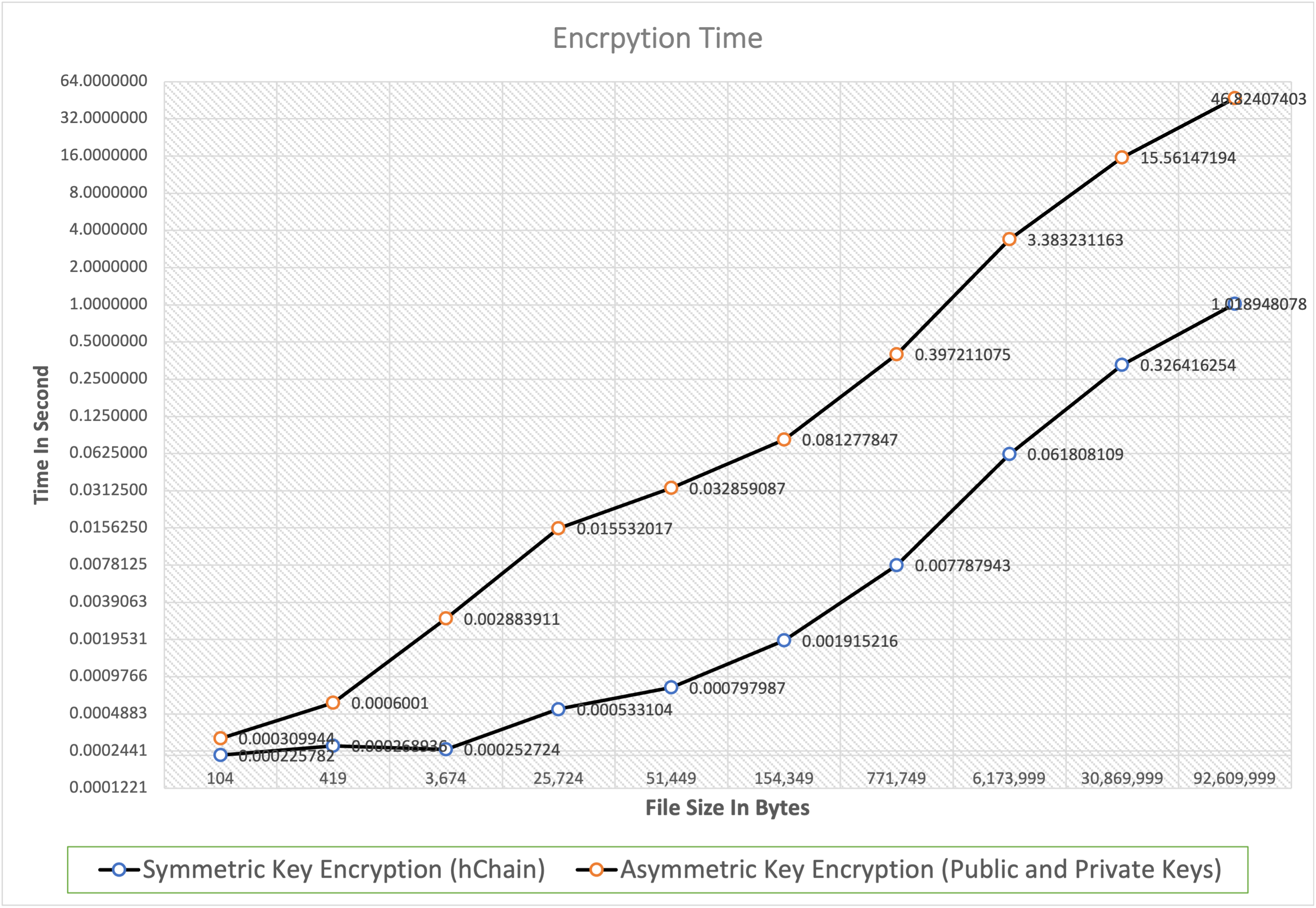}}
	{\includegraphics[width=0.80\textwidth]{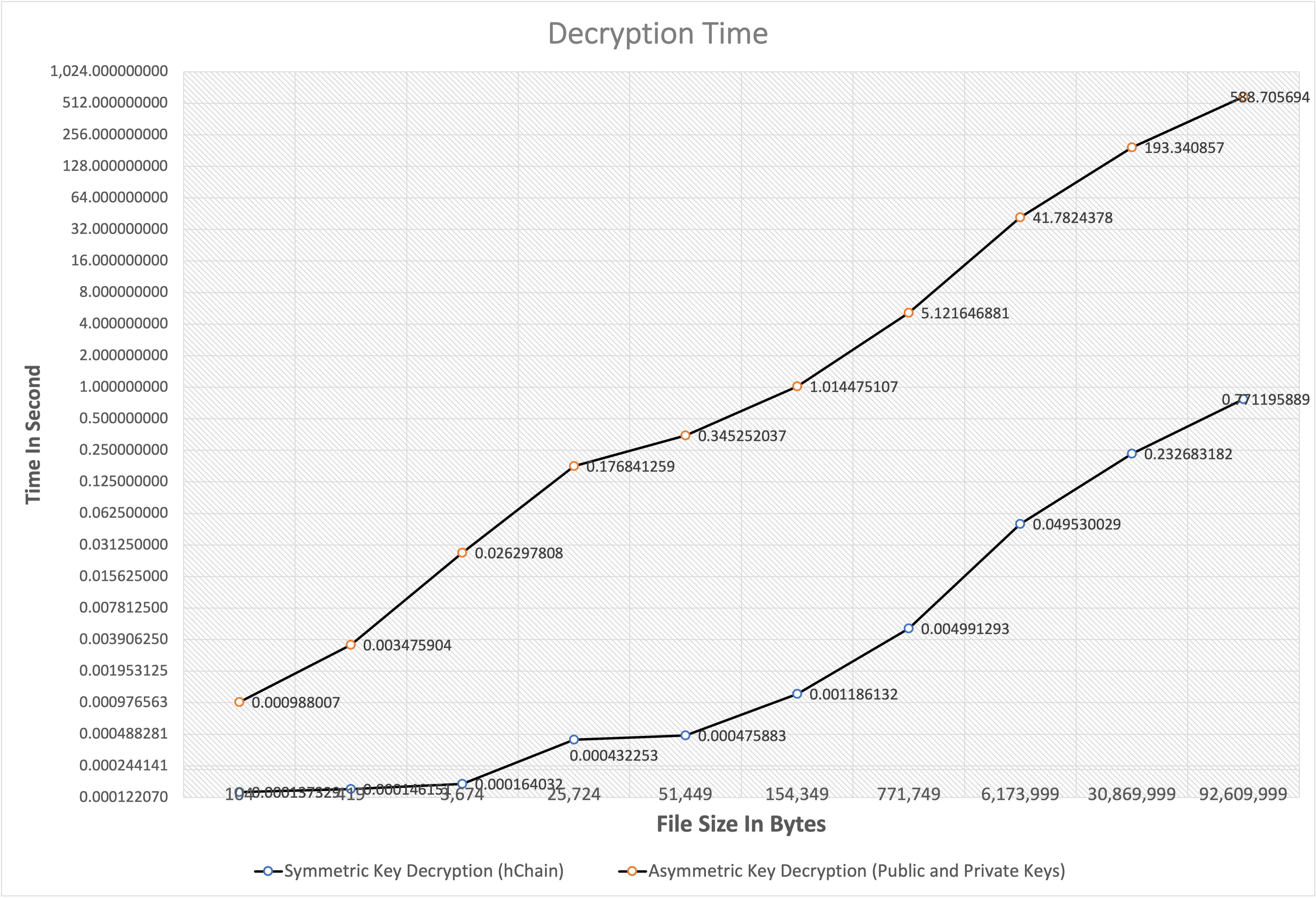}}
	\caption{Encryption And Decryption Comparision Between Symmetric key and Asymmetric key (Public and Private Keys)}
	\label{FIG:Time}   
\end{figure*}

\begin{figure*}[htbp]
	\centering
	{\includegraphics[width=0.80\textwidth]{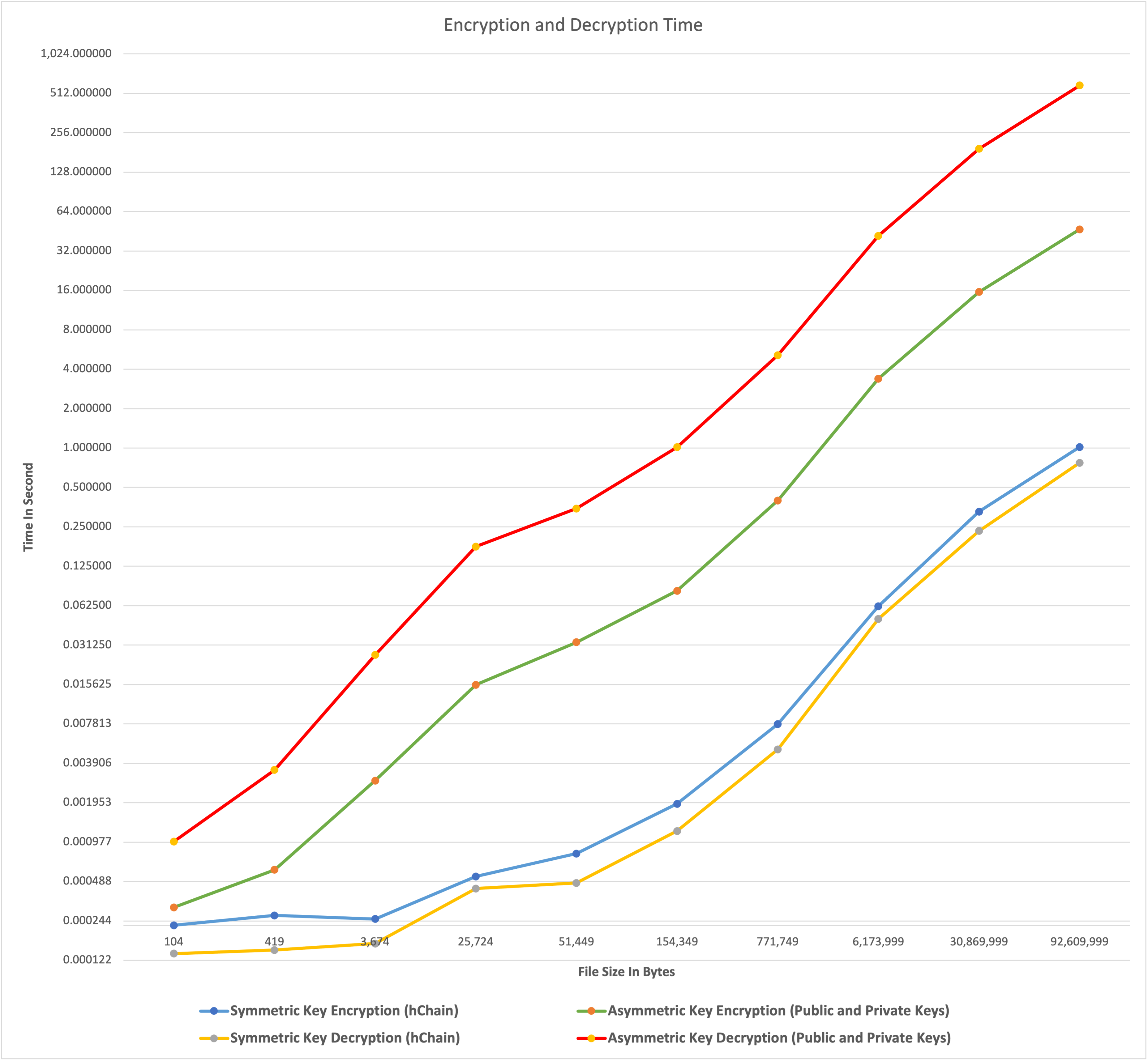}}
	\caption{Encryption And Decryption Comparison}
	\label{FIG:Both}   
\end{figure*}

The time of the encryption The analysis demonstrates the lengths of time required to encrypt all of the data and then decrypt it using both symmetric key encryption and asymmetric key encryption. In order to circumvent latency and the problems that are associated with it in real-time systems, the symmetric key encryption protocol has been introduced at the hChain system. The data is encrypted using an asymmetric key, as described at \cite{Overlay_cloud} . In addition, the key of encrypting the data using the private and public keys takes a significant amount of time, which contributes to an increase in the delay seen in Figures \ref{FIG:Time} and \ref{FIG:Both}. When the amount of data is quite little, the time required varies slightly. On the other hand, when the data is longer than 3000 bytes, the distinction becomes readily apparent. When using symmetric key encryption, both the encrypting and decrypting processes make use of the same key. On the other hand, symmetric key encryption works by encrypting using the public key and decrypting using the private key.

There are a huge difference in time between using the key pairs of blockchain to encrypt and decrypt the data and using symmetric key encryption as it is proposed at hChain.  The decryption time takes around 1000 times of symmetric key encryption when the data is more than 30 million bytes. The encryption time is the same at the latency issue in asymmetric key encryption. It consume much more than symmteric key. The suggested use for encryption is using symmetric key when the system relies on real time. Asymmetric key encryption for very low data is suggested. The way of implementing the encryption and decryption is that encryption the text file size at different size. At asymmetric key encryption, the size is divided into small piece of size due to the data has to be up to the key size. Therefore, it divides the data into key size. 

\section{Conclusion and Future Directions}
\label{sec:conclusions_future}

At every stage of communication, hChain employs a diverse range of cryptographic methods. At the patient level, a symmetric key is combined with a one-way hash function, and a public key infrastructure is utilized to guarantee both data integrity and confidentiality. Moreover, multiple verification factors ensure that data remain unmodified throughout transmission. A verification node, located at the healthcare provider layer, validates both the data and its accompanying signature before creating a transaction on the blockchain. The primary purpose of the smart contract is to manage access controls and grant sensitive information solely to authorized personnel. Blockchain participants are divided into three main membership types—administration, HCP registration, and healthcare provider—each holding distinct permissions that enable them to oversee the entire system. Rather than relying on a centralized infrastructure, blockchain-based EHR solutions provide a secure mechanism for distributing data storage, thereby overcoming the limitations of conventional, center-based healthcare.

Future improvements to this framework will emphasize reducing costs and enhancing portability, thereby facilitating continuous patient monitoring at a high level of security. Furthermore, integrating cloud-based data analysis within the system provides patients with a more comprehensive understanding of their current health status and projected outcomes, potentially helping them avoid adverse consequences related to existing conditions.

\section*{Acknowledgment}
This preprint is based on our conference paper \cite{hchain}.

\bibliographystyle{IEEEtran}
\bibliography{Bibliography_hChain}  

\noindent
\begin{minipage}[t]{0.14\textwidth}
	\vspace{0pt}
	\centering
	\includegraphics[width=0.9in,keepaspectratio]{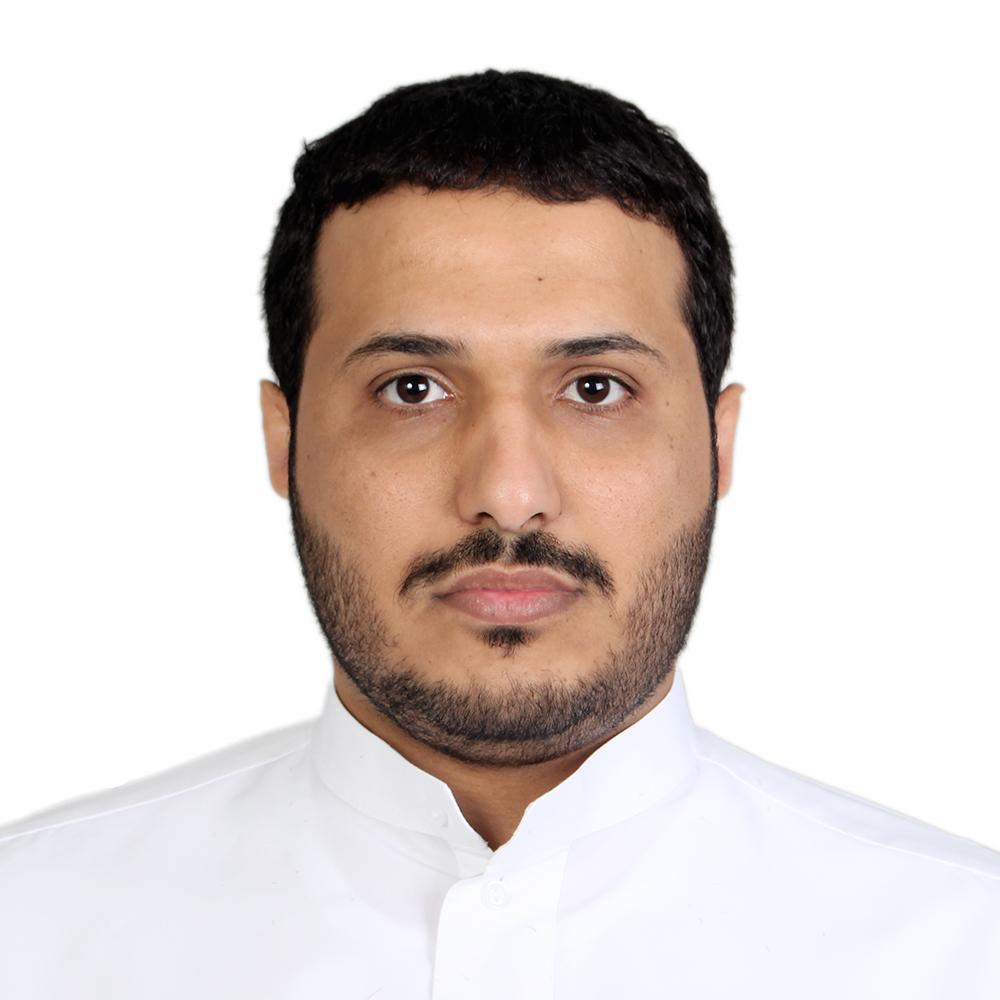}\\
\end{minipage}%
\hfill
\begin{minipage}[t]{0.84\textwidth}
	\vspace{0pt}
	\footnotesize
	Musharraf Alruwaill is a Ph.D. Candidate in the Smart Electronic Systems Laboratory 
	(SESL) at the University of North Texas, under the mentorship of Dr. Saraju Mohanty. 
	He received his Bachelor’s degree from Jouf University and later earned his Master 
	of Science in Computer Science at the University of New Haven. With six publications 
	to his credit, his research interests focus primarily on developing and implementing 
	advanced technologies for smart city applications.
\end{minipage}
\hspace{0.5cm}

\noindent
\begin{minipage}[t]{0.14\textwidth}
	\vspace{0pt}
	\centering
	\includegraphics[width=0.9in,keepaspectratio]{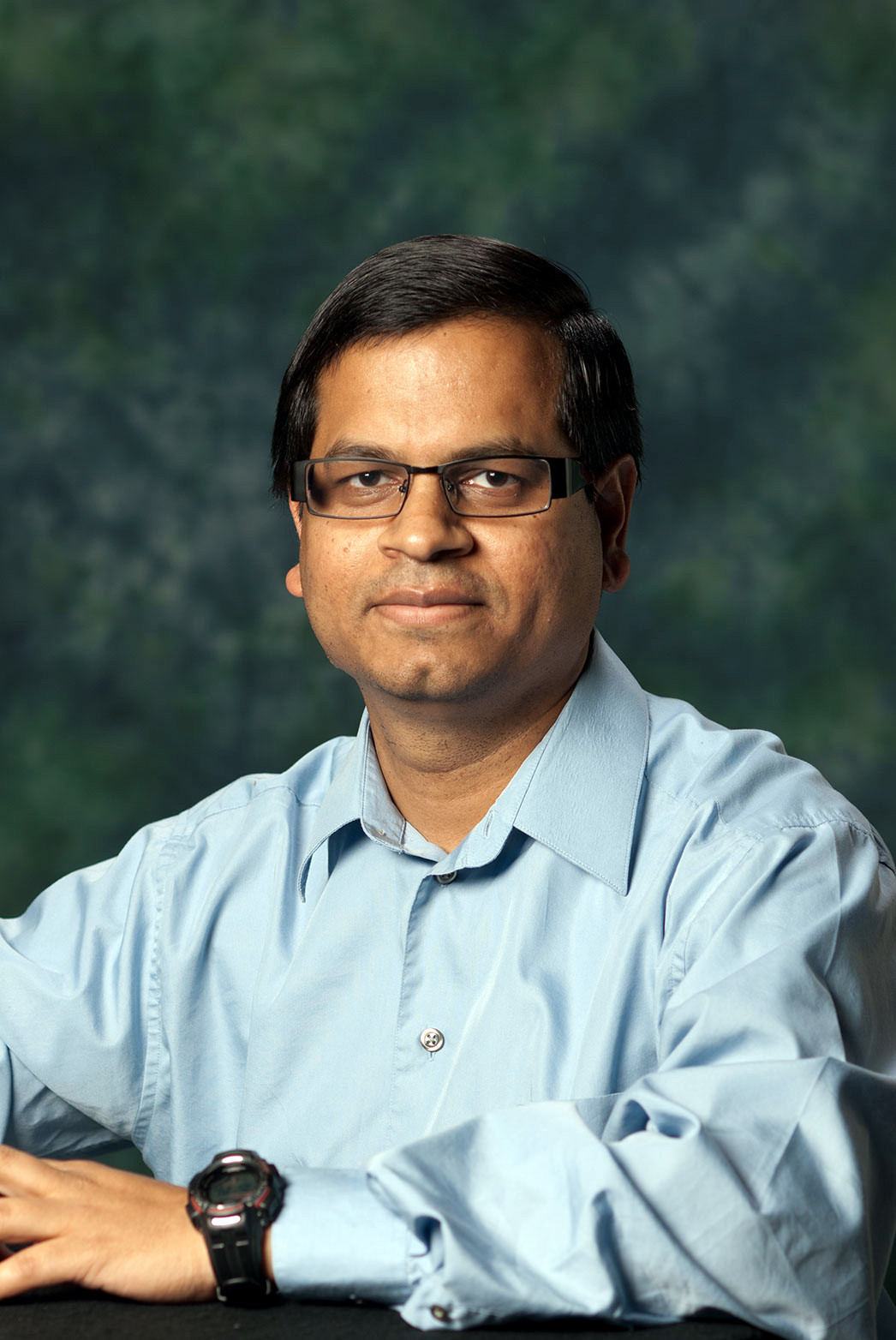}\\
\end{minipage}%
\hfill
\begin{minipage}[t]{0.84\textwidth}
	\vspace{0pt}
	\footnotesize
	Saraju P. Mohanty received the bachelor’s degree (Honors) in electrical engineering 
	from the Orissa University of Agriculture and Technology, Bhubaneswar, in 1995, the master’s degree in Systems 
	Science and Automation from the Indian Institute of Science, Bengaluru, in 1999, and the Ph.D. degree in 
	Computer Science and Engineering from the University of South Florida, Tampa, in 2003. He is a Professor with 
	the University of North Texas. His research is in ``Smart Electronic Systems'', which has been funded by the 
	National Science Foundation, Semiconductor Research Corporation, the U.S. Air Force, NIDILRR, IUSSTF, and 
	Mission Innovation. He has authored 550 research articles, 5 books, and 10 granted or pending patents. His 
	Google Scholar h-index is 62 and i10-index is 298, with 16,000 citations. He is regarded as a visionary 
	researcher on Smart Cities technology, focusing on security, energy-aware design, and AI/ML integration.
	
	He introduced the Secure Digital Camera (SDC) in 2004 with built-in security features using 
	Hardware Assisted Security (HAS) or Security by Design (SbD) principles. He is widely credited as the 
	designer of the first digital watermarking chip in 2004 and the first low-power digital watermarking chip 
	in 2006. Among his accolades are 21 best paper awards, a Fulbright Specialist Award in 2021, the IEEE 
	Consumer Electronics Society Outstanding Service Award in 2020, the IEEE-CS-TCVLSI Distinguished 
	Leadership Award in 2018, and the PROSE Award for Best Textbook in Physical Sciences and Mathematics 
	in 2016. He has delivered 31 keynotes and served on 15 panels at various international conferences.
	
	He has been on the editorial boards of several peer-reviewed international transactions/journals, 
	including \textit{IEEE Transactions on Big Data}, \textit{IEEE Transactions on Computer-Aided Design of 
		Integrated Circuits and Systems}, \textit{IEEE Transactions on Consumer Electronics}, and the 
	\textit{ACM Journal on Emerging Technologies in Computing Systems}. He served as Editor-in-Chief of the 
	\textit{IEEE Consumer Electronics Magazine} from 2016 to 2021, chaired the Technical Committee on Very 
	Large Scale Integration for the IEEE Computer Society from 2014 to 2018, and was on the Board of Governors 
	of the IEEE Consumer Electronics Society from 2019 to 2021. Presently, he serves on the steering, organizing, 
	and program committees of multiple international conferences, including \textit{IEEE International Symposium 
		on Smart Electronic Systems} (IEEE-iSES), \textit{IEEE-CS Symposium on VLSI} (ISVLSI), and \textit{OITS 
		International Conference on Information Technology} (OCIT). Over his career, he has supervised 3 postdoctoral 
	researchers, 18 Ph.D. dissertations, 29 M.S. theses, and 41 undergraduate projects.
\end{minipage}
\hspace{0.5cm}

\noindent
\begin{minipage}[t]{0.14\textwidth}
	\vspace{0pt}
	\centering
	\includegraphics[width=0.9in,keepaspectratio]{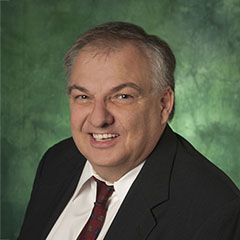}\\
\end{minipage}%
\hfill
\begin{minipage}[t]{0.84\textwidth}
	\vspace{0pt}
	\footnotesize
	Elias Kougianos received a BSEE degree from the University of Patras, Greece in 1985 
	and an MSEE in 1987, an MS in Physics in 1988, and a Ph.D. in Electrical Engineering 
	in 1997, all from Louisiana State University. From 1988 to 1998, he was with Texas 
	Instruments, Inc. in Houston and Dallas, TX. In 1998, he joined Avant! Corp. (now 
	Synopsys) in Phoenix, AZ, as a Senior Applications Engineer, and in 2000 he moved 
	to Cadence Design Systems, Inc. in Dallas, TX, serving as a Senior Architect for 
	Analog/Mixed-Signal Custom IC design.
	
	Since 2004, he has been with the University of North Texas (UNT), where he is 
	currently a Professor in the Department of Electrical Engineering. His research 
	interests include Analog/Mixed-Signal/RF IC design and simulation, as well as 
	the development of VLSI architectures for multimedia applications. He has authored 
	more than 200 peer-reviewed journal and conference publications.
\end{minipage}

\end{document}